\documentclass[aps,prl,reprint]{revtex4-1}
\usepackage{blindtext}
\usepackage{amsmath,amssymb,latexsym}
\usepackage{mathrsfs,graphicx,caption,subcaption}
\usepackage{hyperref}
\usepackage[toc,page]{appendix}

\expandafter\ifx\csname amssym.def\endcsname\relax \else\endinput\fi
\expandafter\edef\csname amssym.def\endcsname{%
       \catcode`\noexpand\@=\the\catcode`\@\space}
\catcode`\@=11
\def\undefine#1{\let#1\undefined}
\def\newsymbol#1#2#3#4#5{\let\next@\relax
 \ifnum#2=\@ne\let\next@\msafam@\else
 \ifnum#2=\tw@\let\next@\msbfam@\fi\fi
 \mathchardef#1="#3\next@#4#5}
\def\mathhexbox@#1#2#3{\relax
 \ifmmode\mathpalette{}{\m@th\mathchar"#1#2#3}%
 \else\leavevmode\hbox{$\m@th\mathchar"#1#2#3$}\fi}
\def\hexnumber@#1{\ifcase#1 0\or 1\or 2\or 3\or 4\or 5\or 6\or 7\or 8\or
 9\or A\or B\or C\or D\or E\or F\fi}

\font\tenmsa=msam10
\font\sevenmsa=msam7
\font\fivemsa=msam5
\newfam\msafam
\textfont\msafam=\tenmsa
\scriptfont\msafam=\sevenmsa
\scriptscriptfont\msafam=\fivemsa
\edef\msafam@{\hexnumber@\msafam}
\mathchardef\dabar@"0\msafam@39
\def\dashrightarrow{\mathrel{\dabar@\dabar@\mathchar"0\msafam@4B}}
\def\dashleftarrow{\mathrel{\mathchar"0\msafam@4C\dabar@\dabar@}}

\def\ulcorner{\delimiter"4\msafam@70\msafam@70 }
\def\urcorner{\delimiter"5\msafam@71\msafam@71 }
\def\llcorner{\delimiter"4\msafam@78\msafam@78 }
\def\lrcorner{\delimiter"5\msafam@79\msafam@79 }
\def\yen{{\mathhexbox@\msafam@55}}
\def\checkmark{{\mathhexbox@\msafam@58}}
\def\circledR{{\mathhexbox@\msafam@72}}
\def\maltese{{\mathhexbox@\msafam@7A}}
\def\circledS{{\mathhexbox@\msafam@73}}

\font\tenmsb=msbm10
\font\sevenmsb=msbm7
\font\fivemsb=msbm5
\newfam\msbfam
\textfont\msbfam=\tenmsb
\scriptfont\msbfam=\sevenmsb
\scriptscriptfont\msbfam=\fivemsb
\edef\msbfam@{\hexnumber@\msbfam}
\def\Bbb#1{{\fam\msbfam\relax#1}}
\def\widehat#1{\setbox\z@\hbox{$\m@th#1$}%
 \ifdim\wd\z@>\tw@ em\mathaccent"0\msbfam@5B{#1}%
 \else\mathaccent"0362{#1}\fi}
\def\widetilde#1{\setbox\z@\hbox{$\m@th#1$}%
 \ifdim\wd\z@>\tw@ em\mathaccent"0\msbfam@5D{#1}%
 \else\mathaccent"0365{#1}\fi}
\font\teneufm=eufm10
\font\seveneufm=eufm7
\font\fiveeufm=eufm5
\newfam\eufmfam
\textfont\eufmfam=\teneufm
\scriptfont\eufmfam=\seveneufm
\scriptscriptfont\eufmfam=\fiveeufm
\def\frak#1{{\fam\eufmfam\relax#1}}

\csname amssym.def\endcsname

\newcommand{\nc}{\newcommand}
\newcommand{\rnc}{\renewcommand}

\nc{\chap}[1]{{\clearpage}%
\begin{center}%
{\noindent\underline{\large\sc #1}}{\addcontentsline{toc}{section}{#1}}%
\end{center}%
{\vspace*{0.3cm}}}

\nc{\subs}[1]{{\vspace*{0.2cm}}%
{\noindent\underline{\small\sc
#1}}%
{\vspace*{0.2cm}}}

\nc{\be}{\begin{equation}}
\nc{\ee}{\end{equation}}
\nc{\bea}{\begin{eqnarray}}
\nc{\eea}{\end{eqnarray}}

\nc{\trac}[2]{{\textstyle\frac{#1}{#2}}}

\nc{\ex}[1]{\mbox{e}^{\,\textstyle#1}}

\nc{\CC}{\Bbb{C}}
\nc{\HH}{\Bbb{H}}
\nc{\PP}{\Bbb{P}}
\nc{\RR}{\Bbb{R}}
\nc{\ZZ}{\Bbb{Z}}
\nc{\II}{\Bbb{I}}
\nc{\EE}{\Bbb{E}}
\nc{\TT}{\Bbb{T}}
\nc{\DD}{\mathrm{I}\!\mathrm{D}}

\rnc{\d}{\delta}
\nc{\eps}{\epsilon}
\nc{\om}{\omega}

\nc{\symx}{\circledS}
\newsymbol\smallsmile 1360
\newsymbol\smallfrown 1361
\nc{\ad}{\mathop{\mbox{ad}}\nolimits}
\nc{\tr}{\mathop{\mbox{tr}}\nolimits}
\nc{\Tr}{\mathop{\mbox{Tr}}\nolimits}
\nc{\Det}{\mathop{\mbox{Det}}\nolimits}
\rnc{\det}{\mathop{\mbox{det}}\nolimits}
\nc{\rk}{\mathop{\mbox{rk}}\nolimits}
\nc{\del}{\partial}
\nc{\diag}{\mathop{\mbox{diag}}\nolimits}
\nc{\ra}{\rightarrow}
\nc{\Ra}{\Rightarrow}
\nc{\LRa}{\Leftrightarrow}
\nc{\lra}{\leftrightarrow}
\nc{\ot}{\otimes}
\rnc{\ss}{\subset}
\nc{\nul}{\noindent\underline}
\nc{\non}{\nonumber\\}
\nc{\mat}[4]{\left(\begin{array}{cc}#1&#2\\#3&#4\end{array}\right)}
\rnc{\lg}{\frak{g}}
\nc{\G}[3]{\Gamma^{#1}_{\;{#2}{#3}}}
\nc{\nam}{\nabla_{\mu}}
\nc{\nan}{\nabla_{\nu}}
\nc{\dx}{\dot{x}}
\nc{\tx}{\tilde{x}}
\nc{\dtx}{\dot{\tilde{x}}}
\nc{\te}{\tilde{e}}
\nc{\dte}{\dot{\tilde{e}}}
\nc{\dxl}{\dot{x}^{\la}}
\nc{\dxm}{\dot{x}^{\mu}}
\nc{\dxn}{\dot{x}^{\nu}}
\nc{\ddx}{\ddot{x}}
\nc{\ddxm}{\ddot{x}^{\mu}}
\nc{\ddxn}{\ddot{x}^{\nu}}
\nc{\dxi}{\dot{\xi}}
\nc{\ddxi}{\ddot{\xi}}
\nc{\lsf}{\ell_s^{\mathrm{eff}}}
\nc{\lpf}{\ell_p^{\mathrm{eff}}}
\nc{\sqg}{\sqrt{g^{11}}}

\nc{\bpm}{\begin{pmatrix}}
\nc{\epm}{\end{pmatrix}}

\nc{\red}[1]{{\color{red}#1}}
\nc{\dd}{\mathrm{d}}

\begin{document}
\title{Wave equations on the linear mass Vaidya metric}
\author{Saeede Nafooshe}
\email{saeede.nafooshe@ung.si}
\author{Martin O'Loughlin}
\affiliation{University of Nova Gorica, Vipavska 13, 5000 Nova Gorica, Slovenia}

\begin{abstract}
We discuss the near singularity region of the linear mass Vaidya metric.
In particular we investigate
the structure in the numerical solutions for the scattering of
scalar and electromagnetic metric perturbations from the singularity. 
In addition to directly integrating the full wave-equation, we use
the symmetry of the metric to reduce the problem to that of an 
ODE. We observe that, around the total evaporation point, quasi-normal like
oscillations appear, indicating that this may be an interesting model
for the description of the end-point of black hole evaporation. 
\end{abstract}

\maketitle

\ifpdf
    \graphicspath{{6/figures/PNG/}{6/figures/PDF/}{6/figures/}}
\else
    \graphicspath{{6/figures/EPS/}{6/figures/}}
\fi  

\section{Introduction}
The Vaidya metric is a useful solution to Einstein's equations with 
a stress-energy tensor that corresponds to an outgoing, spherically 
symmetric flux of radiation \cite{vaidya}. It has been used as a model
for the metric outside stars that includes the back-reaction of the space-time
to the stars radiation, and also as a model for various studies of both black 
hole formation and evaporation \cite{waugh862,bicak97,hiscock82,kuroda,
balbinot,abdalla,bicak03,ghosh,harko,girotto,kawai,fayos10,farley}.
The linear mass Vaidya metric is a special class of Vaidya metrics over which 
one has a certain degree of analytic control, in particular as a consequence
of the additional homothety symmetry that these metric possess. For a restricted
range of parameters in the outgoing Vaidya metric with linear mass 
the metrics contain a null singularity that vanishes at a point internal to the 
space-time and thus is an ideal exact candidate for a model of a decaying 
black hole. 

In a previous paper \cite{OLoughlin:2013aa} a particular scenario
was introduced and an initial study of the behaviour of metric perturbations
was presented in support of this model. 
The out-going Vaidya metric with monotonically decreasing linear mass function 
can resemble a realistic situation for the final phase of black hole 
evaporation.
In this paper we study in more detail the electromagnetic and scalar 
perturbations of the out-going linear mass Vaidya metric in this context,
in particular to study the perturbation equations of this dynamical 
space-time results looking for a quasi-normal (QN) like ringing. 
Such results would give support to the claim that this metric 
is black-hole like around the vanishing point of the singularity and thus is
suitable to be considered as the transitional state between an adiabatically 
evaporating Schwarzschild black hole at the end stage of its life and 
Minkowski space-time. 

Quasi-normal modes (QNMs) \cite{qnmbs1,qnmbs2} for time-dependent backgrounds 
have been investigated in 
\cite{Hod:2002gb,Xue:2003vs,Shao:2004ws} and in particular for in-going Vaidya 
metric in \cite{invai1,invai2}. The general shape of the oscillations for 
dynamical backgrounds like Vaidya is different from that of the stationary 
ones like Schwarzschild. In the stationary adiabatic regime the real part 
of QNMs changes inversely with the mass function. For dynamical 
backgrounds where the mass changes with time, the period of the oscillation 
will 
also change, thus the shape of the waveform includes oscillations with 
varying periods. 
The power law fall off of the tail of QNMs originally calculated by Price 
\cite{price1972} for stationary space-time is also different for dynamical 
backgrounds \cite{Hod:2002gb, Hod:2009my}. Numerical errors in the 
investigation of tail phenomena in dynamical background are unavoidable so 
to have a better picture of this phenomena one should also more analytic
methodes if they are available. 

In this paper we use both numerical and analytical methods to study the 
response of the out-going Vaidya background to the electromagnetic and scalar 
perturbation. We first write the perturbation equations in double null 
coordinates \cite{Waugh:1986aa} and then we solve the partial differential 
equations (PDE) numerically. To provide an alternative, more analytic approach
we then use the homothety symmetry 
of the linear mass vaidya metric to reduce the problem to that of an ordinary
differential equation and comment on the results.

\section{Outgoing Vaidya Space-Time}
The Vaidya metrics \cite{vaidya} are exact solutions of the Einstein 
equations. In radiation coordinates $(w, r, \theta, \phi)$ this metric 
has the form
\begin{equation}
ds^{2}=-(1-\frac{2 m(w)}{r})dw^{2}+2c dw dr +r^{2} d\Omega^{2}, 
\end{equation}
where $c=1,-1$ respectively corresponds to ingoing and outgoing radial flow,
$w = t + cr$
and $m(w)$ is a monotonic mass function. In the presence of spherical symmetry 
this mass function can be the measure of the amount of energy within a sphere 
with radius $r$ at a time $t$ \cite{Zannias:1990aa, Nielsen:2008kd}.

The causal and singularity structure of this space-time can change 
significantly with the choice for the mass function. 
For constant mass this solution reduces to the Schwarzschild solution in 
ingoing or outgoing Eddington-Finkelstein coordinates.
The ingoing Vaidya  metric describes collapsing 
null dust \cite{Gao:2005yq}. The outgoing Vaidya space-time 
\begin{equation}\label{outgoingv}
\begin{split}
&ds^{2}=-f(u,r)du^{2}-2 du dr +r^{2} d\Omega^{2} \\
&f(u,r)=(1-\frac{2 m(u)}{r})
\end{split}
\end{equation}
describes the evolution of a radiating star or black hole, where $m(u)$ is 
the mass function of retarded time $u$ that labels the outgoing radial null 
geodesics. In the following we will restrict our analysis to the outgoing
case as we are interested in the final stages of black hole evaporation. 

The only non-vanishing component of the Einstein tensor is 
\begin{equation}\label{nonvanish}
G_{uu}=-(\frac{2}{r^{2}})\frac{d m(u)}{du},
\end{equation}
and the stress-energy tensor that leads to this solution is
\begin{equation}\label{setvaidya}
T_{\alpha \beta}=-\frac{1}{4 \pi r^{2}}\frac{d m(u)}{du} k_{\alpha}k_{\beta}
\end{equation}
where $k_{\alpha}$ is tangent to radial outgoing null geodesic, 
$k_{\alpha}k^{\alpha}=0$. This stress-energy tensor describes a pressure less 
fluid with energy density $\rho =-\frac{d m(u)}{du}4 \pi r^{2} $ moving with 
four-velocity $k_{\alpha}=\delta_{\alpha}^{u}$ (such a fluid is called 
``null dust''). To satisfy the null energy condition for which $\rho \geq 0$, 
the mass function $m(u)$ must be a decreasing function of increasing retarded 
time, namely $\frac{d m(u)}{du} <0$, which means that the mass function 
decreases in response to the outflow of radiation as one would expect for the 
evolution of a radiating star or an evaporating black hole.  
For our analysis we will choose the linear mass function $m(u)=-\mu u$. This 
choice of mass function will enable us to study the possible evolution of 
the space-time around the end point of black hole evaporation.

In addition to the spherical symmetry of this space-time
\eqref{outgoingv} it is also homothetic in the case that the 
mass function is linear. The space-time 
possesses a conformal Killing vector $K$\cite{Hiscock:1982pa} 
\begin{equation}\label{Killingeq}
K_{\mu;\nu}+K_{\nu;\mu}=2\rho g_{\nu \mu}
\end{equation}
where $\rho$ is a constant, indicating that this is actually a homothety 
symmetry. Homothety means that the metric with linear mass 
function scales upon a scaling of the coordinates by an overall factor
\begin{equation}\label{scalein}
(u,r) \rightarrow (\zeta u,\zeta r)\quad\quad \Rightarrow \quad\quad 
ds^{2} \rightarrow \zeta^{2} ds^{2},
\end{equation}
for any real $\zeta$.

\subsection{The conformal structure of linear-mass Vaidya space-times}
\label{confdia}
\begin{figure}[h!]
\center{\includegraphics[width=5cm]{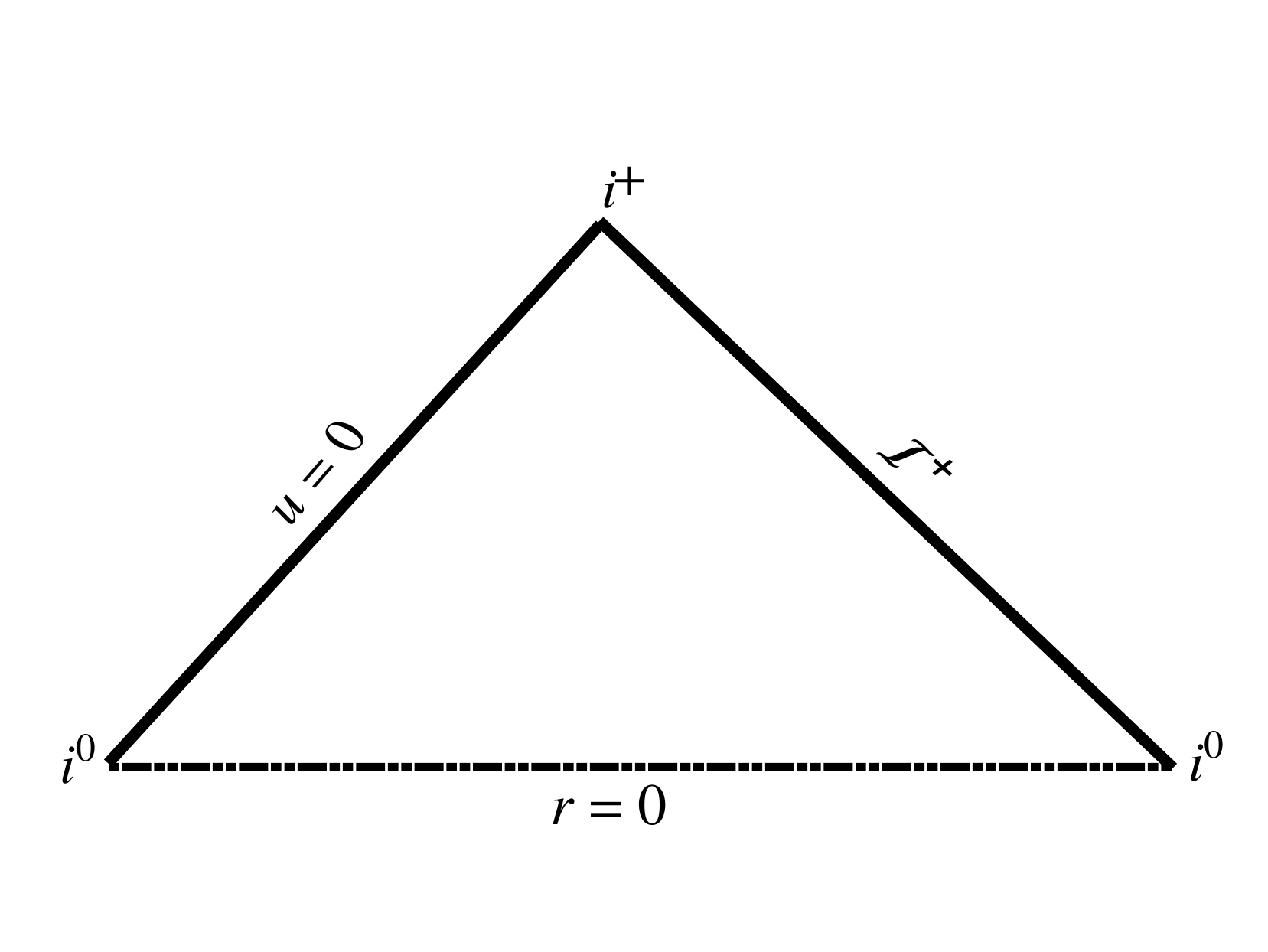}}
\caption{Conformal diagram for outgoing Vaidya with linear mass function for 
$\mu>1/16$, dot-dashed line represents $r=0$ singularity.}
\label{penrose}
\end{figure}

In general the choice of mass function in Vaidya space-time determines its
global and local structure and singularities. 
Here we will consider only the case of 
a linear mass function $m(u)=-\mu u$ and the conformal structure of the 
space-time varies with $\mu$ \cite{Waugh:1986aa} in the following way. For 
$\mu>1/16$ the conformal diagram is displayed in figure \eqref{penrose}. 
\begin{figure}[h!]
\center{\includegraphics[width=7cm]{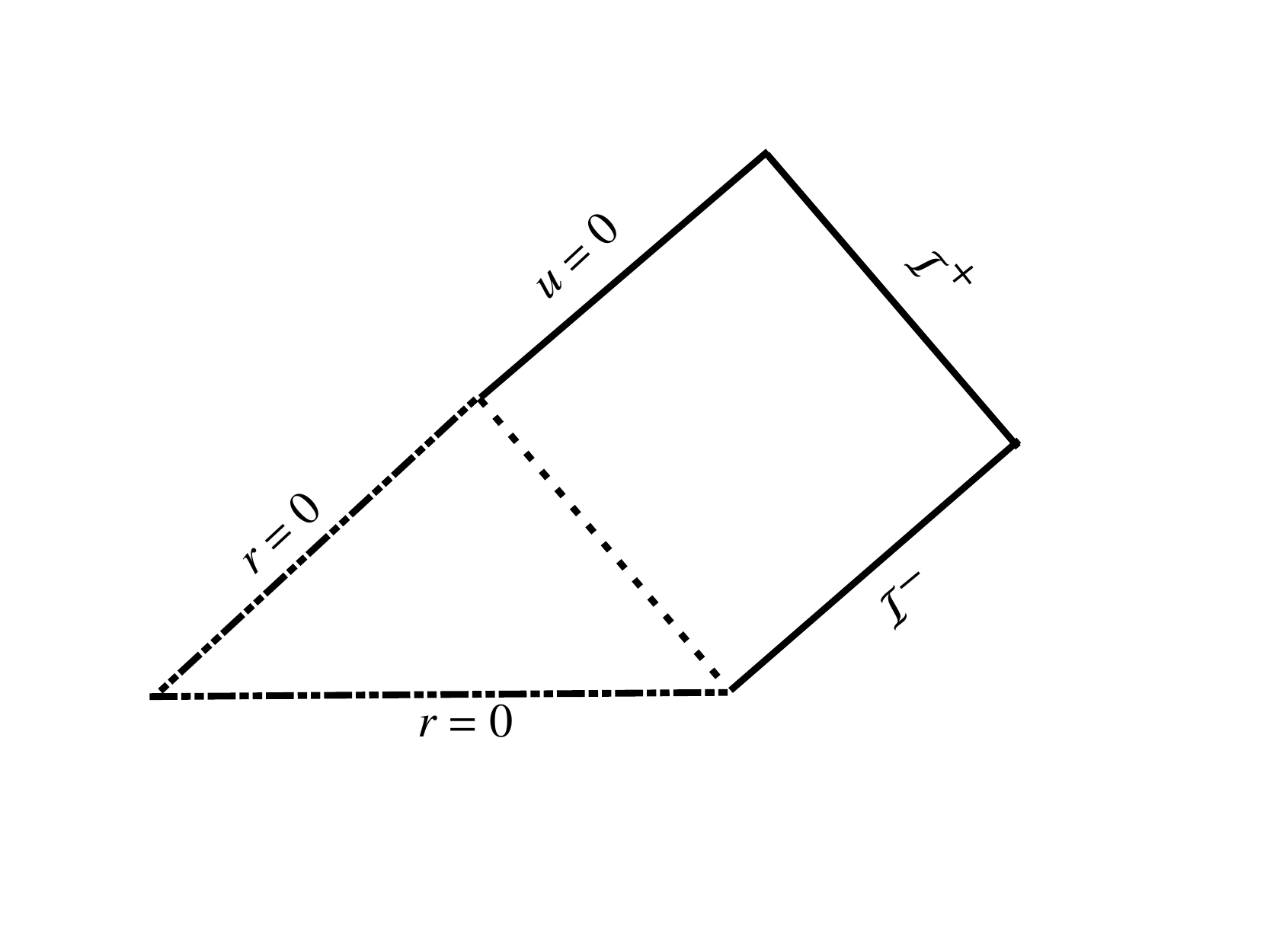}}
\caption{Conformal diagram for outgoing Vaidya with linear mass function for 
$\mu=1/16$, dot-dashed lines represent $r=0$ singularities.}
\label{penrose2}
\end{figure}
\begin{figure}[h!]
\center{\includegraphics[width=7cm]{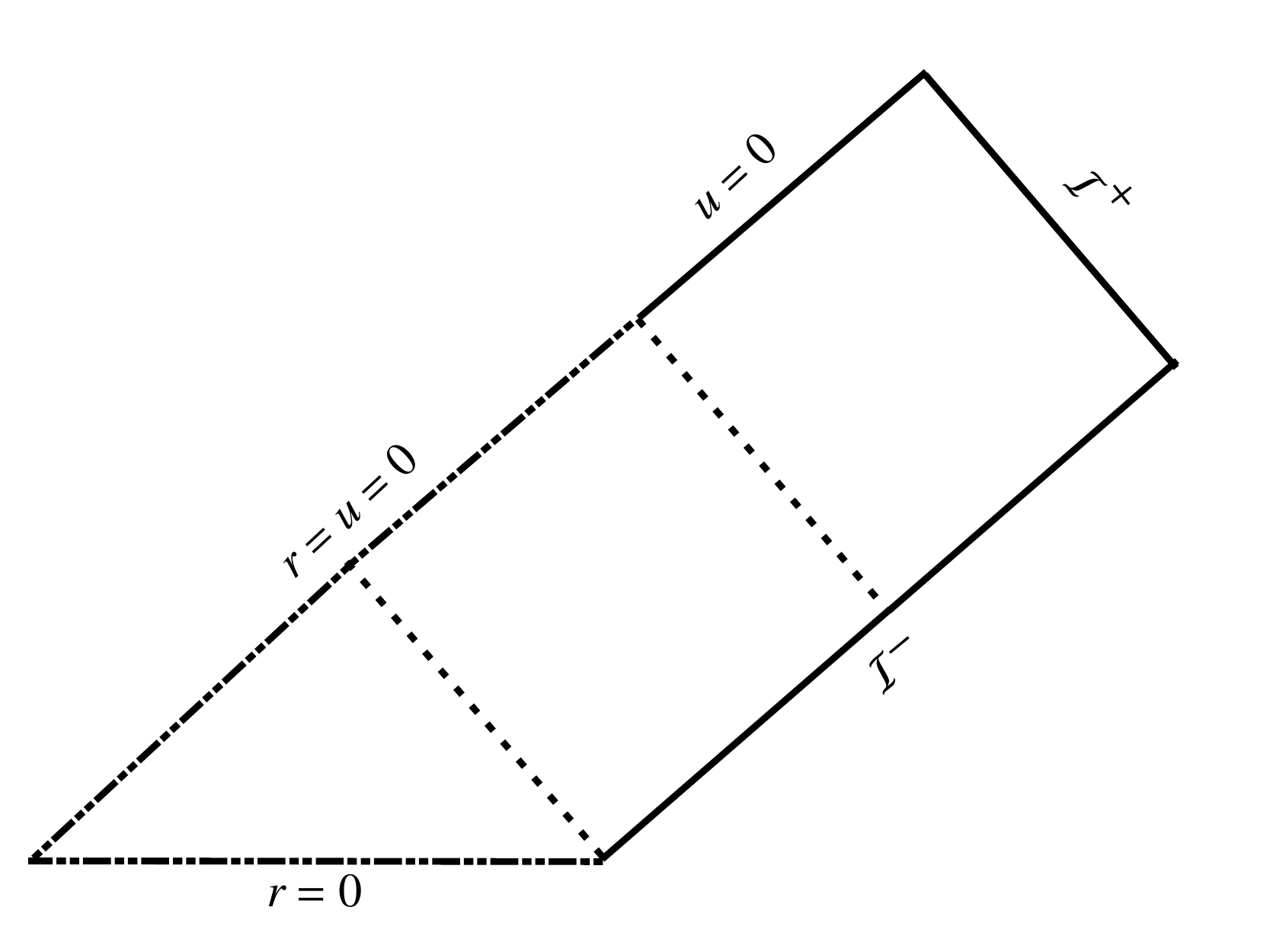}}
\caption{Conformal diagram for outgoing Vaidya with linear mass function for 
$\mu<1/16$, dot-dashed lines represent $r=0$ singularities.}
\label{penrose3}
\end{figure}
The dot-dashed line shows the singularity at $r=0$ for $u<0$. The next case is 
$\mu=1/16$ which is represented in figure \eqref{penrose2}. In the last case 
in figure \eqref{penrose3} the conformal 
diagram for $\mu<1/16$ is shown. In this case the $u=0$ boundary to the 
future of the 
endpoint of the $r=u=0$ singularity is special in that the space-time there
approaches that of Minkowski space. Indeed it has been shown in 
\cite{Unruh:1985aa} that one can continuously attach the metric along 
this part of the $u=0$ hypersurface to Minkowski space without 
introducing curvature singularities.

\begin{figure}[h!]
\center{\includegraphics[width=10cm]{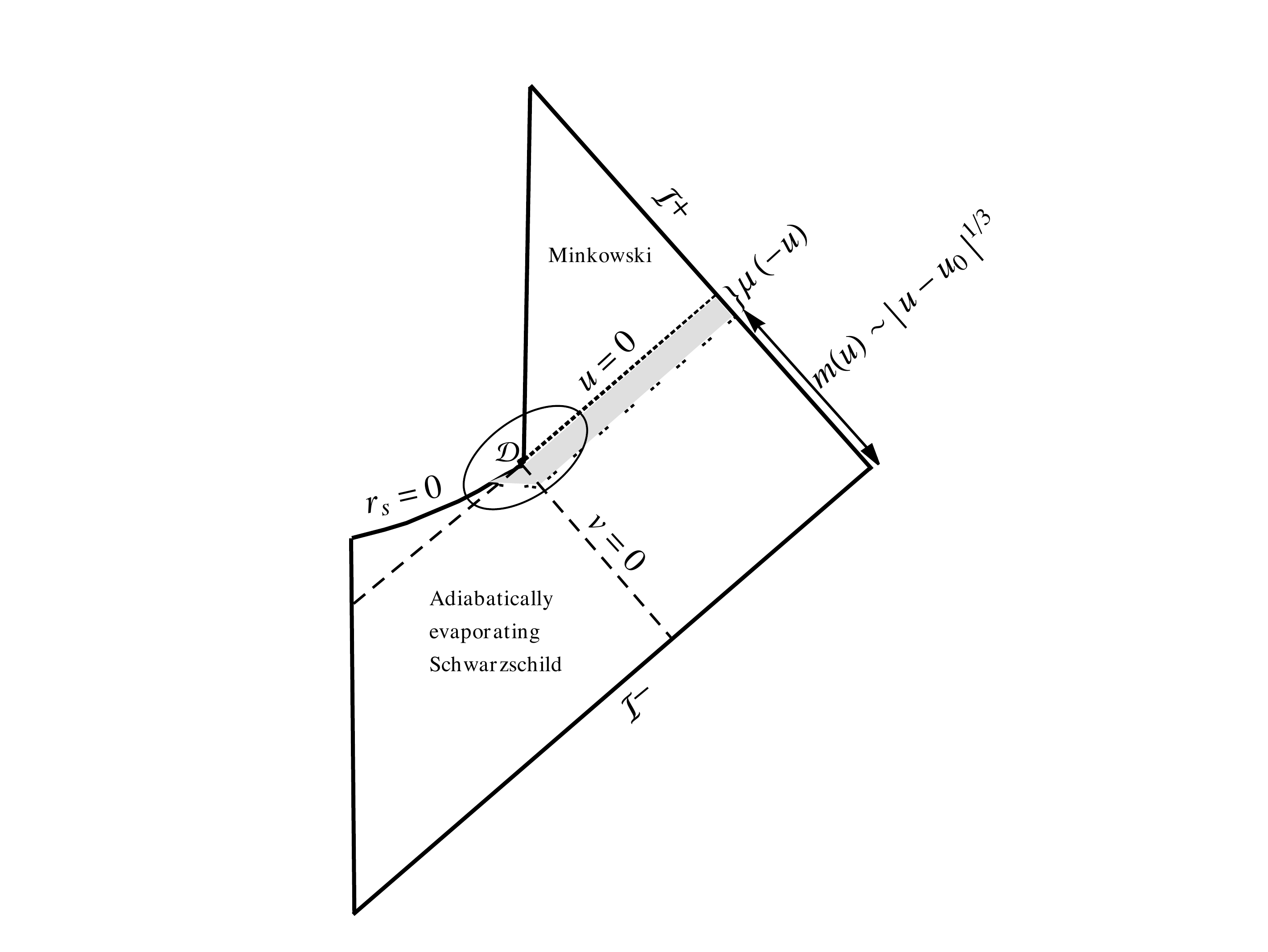}}
\caption{Conformal diagram for the evaporation of a Schwarzschild black 
hole with linear Vaidya at the final stage of evaporation and 
a future Minkowski region.}
\label{conformalov}
\end{figure}
Considering the outgoing Vaidya metric with linear mass function and with 
$\mu<1/16$, a new model for the final fate of a black hole at the end of its 
evaporation has been proposed in \cite{OLoughlin:2013aa}.
This space-time can be divided into three different regions characterized by a
transition time $u_t$ and illustrated in figure 4: an adiabatic Schwarzschild
region for all $v$ with $u<u_t$ with $m(u)\sim |u-u_0|^{1/3}$ and also 
most of the region $v<0$; a Vaidya region 
with linear mass function for $u_t<u<0$, $v\geq 0$; a Minkowski space-time 
region for $u>0$, $v>0$. In this model the linear mass function is used  when 
the mass of black hole becomes Planckian. 

\subsection{Vaidya in Double Null Coordinates}
As our purpose is to study wave-equations on the outgoing Vaidya 
space-time, it is very useful to introduce the double null coordinate 
\cite{Waugh:1986aa} for which both semi-analytical and numerical calculations 
can be performed. In these coordinates $(u,\theta,\phi,v)$ the general 
form of the metric is 
\begin{equation}\label{vdouble}
ds^{2}=-2 f(u,v)du dv+r^{2}(u,v) d\Omega^{2} 
\end{equation}
For the outgoing metric, the energy momentum tensor has the form
\begin{equation}\label{tdouble}
 T_{\mu\nu}= \frac{\mu}{4 \pi r(u,v)^{2}} (\delta_{\mu}^{u})(\delta_{\nu}^{u})
\end{equation}
Considering the linear mass function with $\mu<1/16$ and introducing 
$\Delta=\sqrt{1-16\mu}$, $f(u,v)$ is 
\begin{equation}\label{fdouble}
 f(u,v) = \frac{1+\Delta}{2\Delta r(u,v)} (r(u,v)+u(1-\Delta)/4)^{2/(1+\Delta)}, 
\end{equation}
where $r(u,v)$ can be derived by solving this equation
\begin{equation}\label{rdouble}
\begin{split}
& \left(\frac{v}{|u|^{2\Delta/(1+\Delta)}}\right)^{1+\Delta} 
\left(\frac{r(u,v)}{|u|}-\frac{1-\Delta}{4}\right)^{1-\Delta} \\
&= \left(\frac{r(u,v)}{|u|}-\frac{1+\Delta}{4}\right)^{1+\Delta}.
\end{split}
\end{equation}
The function $r(u,v)$ can be found exactly for $\Delta = 3/5,1/2,1/3,1/5,1/7$
and the explicit solutions have been given in \cite{OLoughlin:2013aa}.

\section{Vaidya Potential} 

In general QNMs are are found as decaying oscillations in 
the metric perturbation close to the horizon of a black hole.
The frequencies of these modes generally have a complex form of which the real 
part represent the oscillation frequency and the imaginary part represents the 
damping of the oscillation. QNMs can be calculated for both stationary and 
time dependent background and they are black holes fingerprints. 
The evolution of the 
response of the black hole to perturbations can be divided in three stages: 
first an initial wave burst in a relatively short time by the source off 
perturbation, then the ``ringing radiation'' which is caused by the damped 
oscillations of QNMs that are excited by the source of perturbation and 
finally a power law tail suppression of QNMs at very late time due to the 
scattering of the wave by the effective potential. 

In order to study possible QN like modes of Vaidya space-time, 
we need to 
study the wave equations for perturbations of the space-time metric 
\cite{ReggeWheeler} which are naturally divided into
scalar, electromagnetic and tensorial modes,
\begin{equation}\label{kg} 
\frac{\partial^{2}\psi}{\partial u \partial v}+ W(u,v) f(u,v) \psi=0
\end{equation}
where $W(u,v)$ is given by
\begin{equation}\label{potential} 
W(u,v)=\frac{\ell (\ell+1)}{2 r^{2}(u,v)} + \sigma \frac{m(u)}{r^{3}(u,v)} 
\end{equation}
and where $\sigma = 1$ and $\sigma = 0$ correspond, respectively, to the scalar 
and electromagnetic perturbations on which we will focus the current study. 
From here on, for calculational convenience, we extend the linear 
mass function $m(u)=-\mu u$ to all values of $u<0$ and not just for the 
$u_t<u<0$ as was shown in figure \eqref{conformalov}. Equation 
\eqref{kg} describes wave propagation in the Vaidya background and 
$f(u,v) W(u,v)$ is the effective potential which describes how 
fields are scattered by the geometry. It is clear 
that this potential depends on the black hole geometry and also on the 
spin of the perturbation under consideration.

\subsection{Integrating the PDE}

We proceed by using the numerical integration technique for 
the calculation of QNMs
originally proposed and developed in \cite{Gundlach:1994}. 
In the present context this equation was already 
studied for the special case of electromagnetic perturbations with $\ell = 1$
in \cite{OLoughlin:2013aa} where it is was observed 
that an initially ingoing gaussian wave-packet coming in from ${\mathcal I}^-$ 
with centre at small negative $\bar{v}$ appears to develop a QN like 
ringing as it evolves towards $\bar{u}\to 0$.The numerical integration was 
carried out by sending in the direction of increasing $u$ a gaussian wave 
localized around $v_c<0$. 

\begin{figure}[h!]
        \centering
        \begin{subfigure}{0.4\textwidth}
                \includegraphics[width=\textwidth]{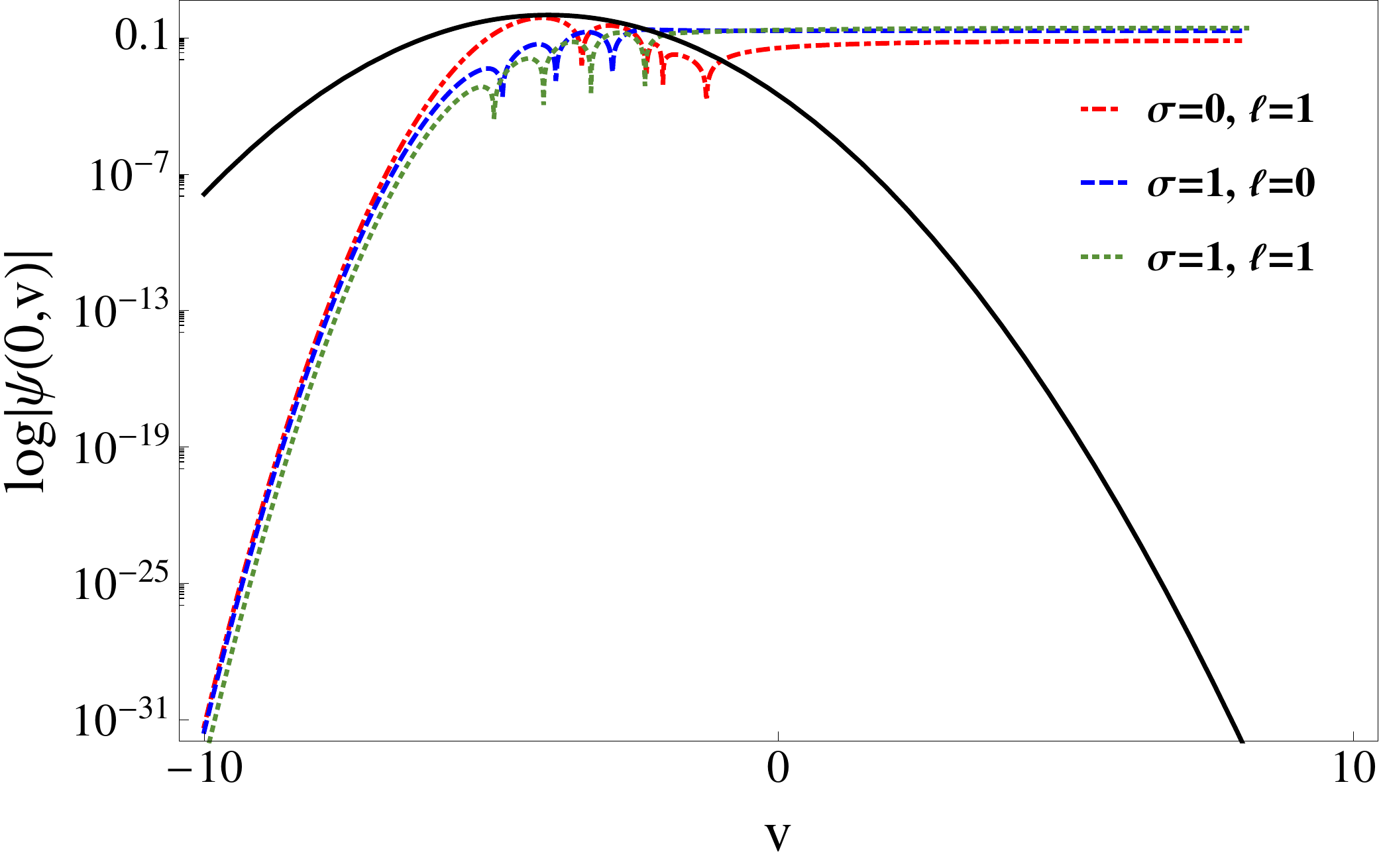}
                \caption{$w=1$, $v_c=-4$}
                \label{12ev4w1}
        \end{subfigure}\hfill%
        \begin{subfigure}{0.4\textwidth}
                \includegraphics[width=\textwidth]{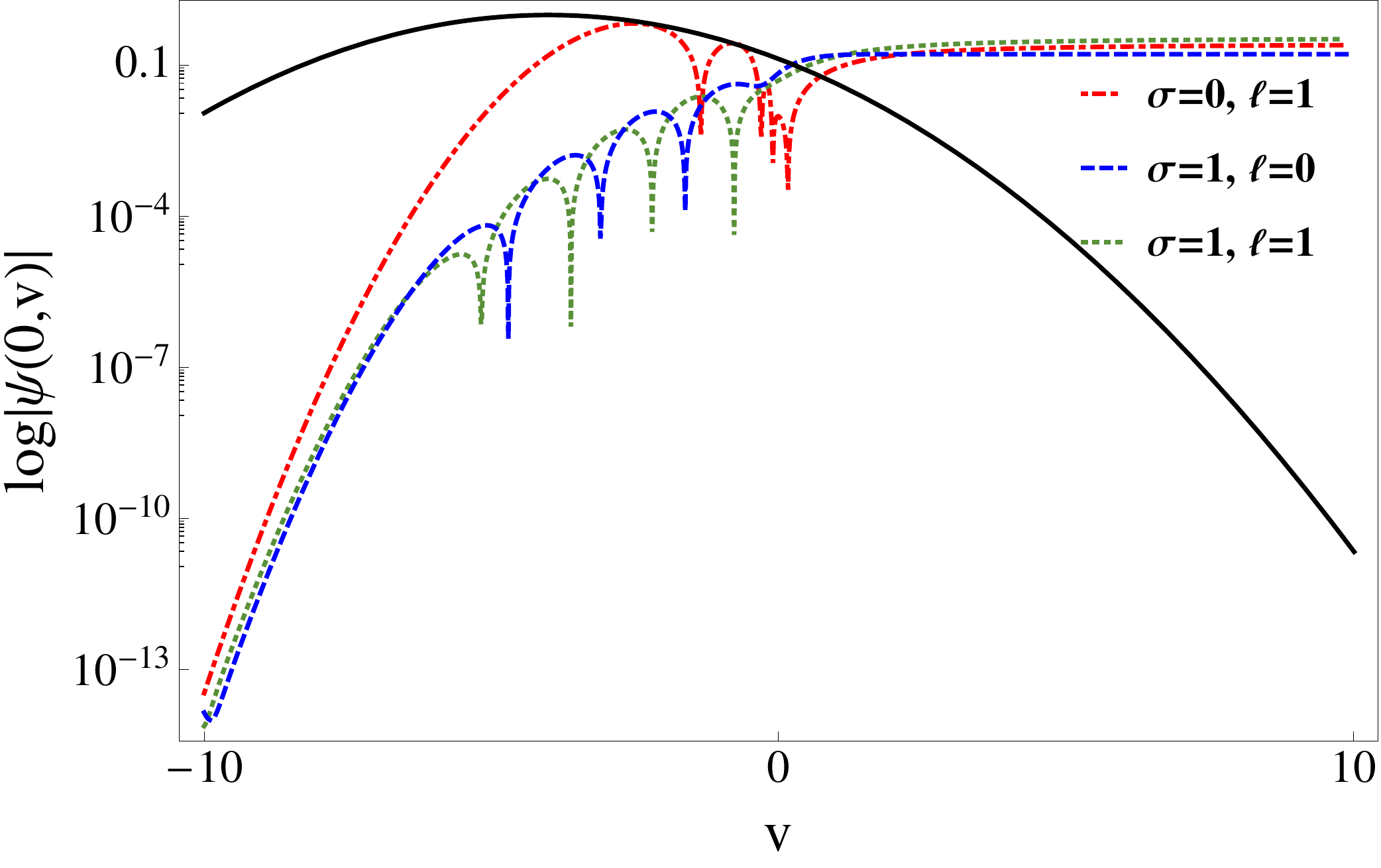}
                \caption{$w=2$, $v_c=-4$}
                \label{12ev4w2}
        \end{subfigure}\hfill%
       \begin{subfigure}{0.4\textwidth}
                \includegraphics[width=\textwidth]{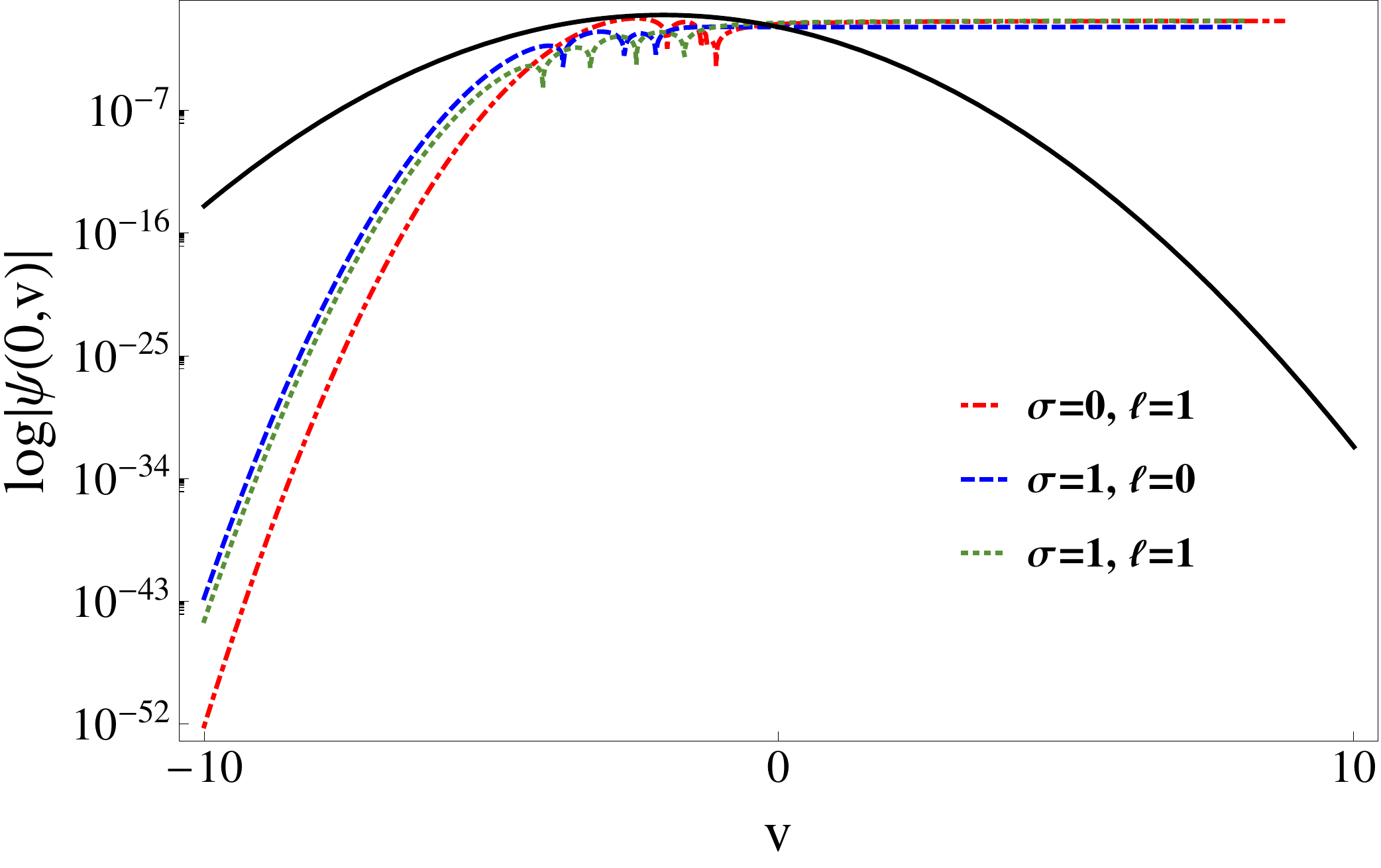}
                \caption{$w=1$, $v_c=-2$}
                \label{12ev2w1}
        \end{subfigure}\hfill%
        \caption{Time profile of the response of outgoing Vaidya space-time to the electromagnetic and scalar perturbations for $\Delta=1/2$ and $\ell=0, \ell=1$ for different values of the initial data. The solid curve indicate the Gaussian function that 
has been used as initial data where $w$ is the width and $v_c$ marks the center of the Gaussian. }\label{12e}
\end{figure} 

In this paper, in addition to the calculation for the electromagnetic field
we also present the numerical integration to obtain the time 
profile of the perturbed outgoing Vaidya for both electromagnetic $\sigma=0$ 
and scalar perturbations $\sigma=1$ and for different angular momentum values. 
Some selected results for the evolution of the ingoing 
wave are presented in figure \eqref{12e}. 
In these figures the results of the integration with $\Delta=1/2$ are 
displayed. Similar results can also be obtained for other values of $\Delta$.
The initial conditions were a gaussian wave form in $v$ with centre at $v=v_c$
at $u=u_{0}=-40$ and with varying widths. 
One can see that in particular there is a ringing of varying period 
for $\bar{v}\lesssim 0$. The ringing dies out rapidly and is not present
for $v>0$ in line with the fact that the ``Planckian'' black hole has 
vanished. The general form 
of these oscillations doesn't change for different values of the initial 
gaussian, though their detailed structure does. This indicates that there 
are not true QNMs at particular discrete frequencies in 
contrast to what one finds for the Schwarzschild black hole.

These results are in line with earlier studies of QNMs for 
dynamical backgrounds \cite{Xue:2003vs} where it was been pointed out 
that when the black hole mass decreases with time the oscillation 
period becomes shorter in contrast to the constant frequency QNMs of the 
Schwarzschild black hole.
These solutions show a constant tail after few oscillations for large values 
of $v>0$ however we will see in the next section that as a consequence 
of the homothety of the metric and the initial conditions 
that the $|\psi|\ra \text{const}$ behavior at 
large positive $v$ is most likely a consequence of numerical errors. 
In \cite{Hod:2002gb, Hod:2009my} has been shown that the there is time window 
between the dominant period of QN ringing and the tail of these 
modes. In effect the tail behavior with a pure power law decay is only 
expected at infinitely late times. In practice the numerical integration 
is for a finite time interval and this causes an inherent error in the 
behavior of the tail. 

In the next subsection we will show that, as a consequence of the 
scaling symmetry, the wave-equation can be separated, thus reducing the 
problem to that of an ordinary differential equation. We will also see 
from the separation ansatz that evolution is essentially a frequency 
dependent rescaling of the modes that are used to construct the initial 
Gaussian profile.

\subsection{Reduction to an ODE}
The main purpose of the current research was to present the wave-profiles that 
one can obtain from the numerical mesh integration method for different 
initial conditions and fields, as carried out in the previous section, 
and then to compare them with the individual mode solutions 
that we will obtain below via a semi-analytic 
method that takes advantage of the scaling symmetry of the space-time and 
equations. We will now look at individual modes of the wave-function that we 
obtain by using the homothety symmetry of the equations to carry
out a separation of variables in the differential equation \eqref{kg}.

\begin{figure}[htbp]
\begin{center}
 \includegraphics[width=7cm]{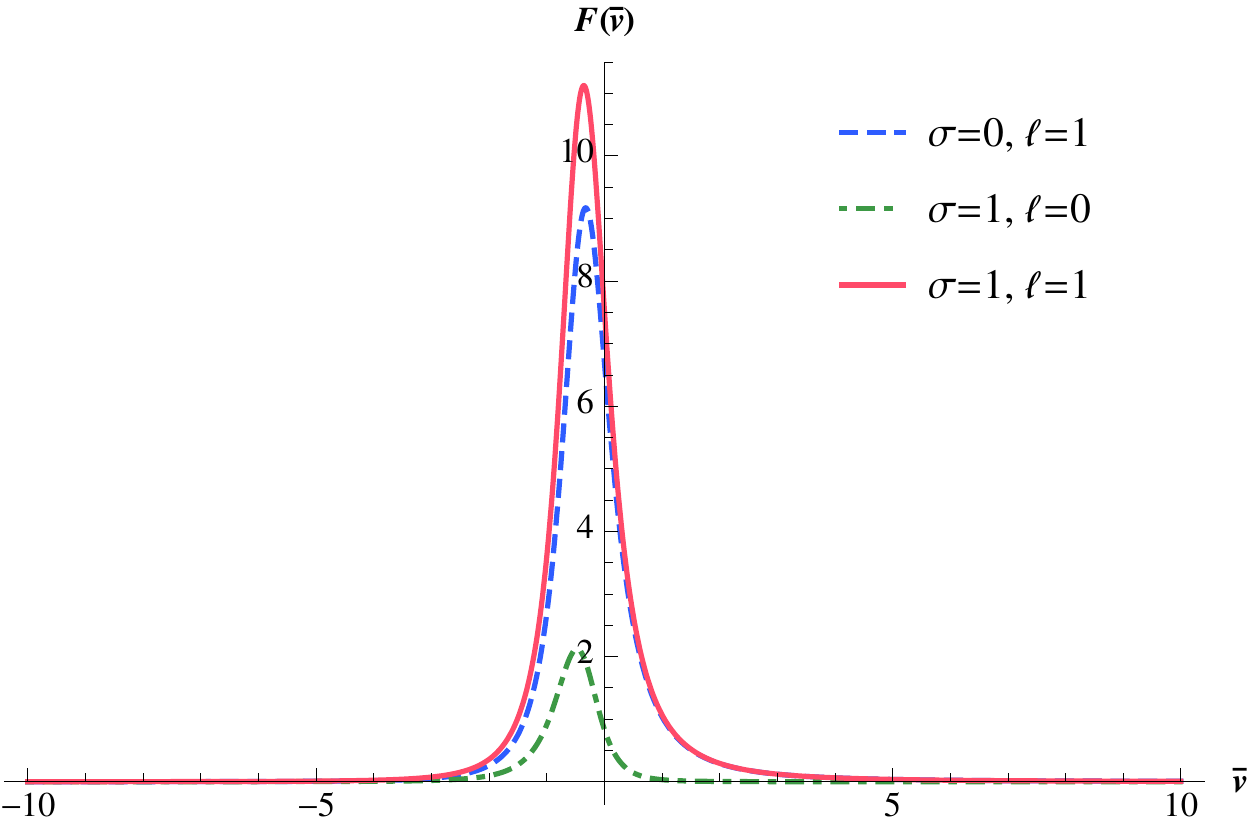}
\caption{(a) $F(\bar{v})$ for $\Delta=1/2$ and $\sigma=0,1$, for $3$ different 
values of angular momentum}\label{fvbard12}    
\end{center}
\end{figure}

The homothety symmetry of this space-time suggests that we change the 
variable as follows 
\begin{equation}\label{cvaribale}
\bar{u} = -u = |u|,   \quad\quad  \bar{v} = v (-u)^{-2 \Delta/(1+ \Delta)},
\end{equation}
giving (from \eqref{rdouble})
\begin{equation}\label{g}
r = r(u,v) = |u| g(v/|u|^{2\Delta/(1+\Delta)}).
\end{equation}
Applying these changes to equation \eqref{kg}

together with the ansatz
\begin{equation}\label{kap}
\psi_\lambda(\bar{u},\bar{v}) = \bar{u}^{\lambda} V_\lambda(\bar{v}),
\end{equation}
we obtain the following differential equation
\begin{equation}\label{kg4}
\bar{v}\frac{\partial^{2} V(\bar{v})}{\partial \bar{v}^{2}} + ( 1 -\kappa )
\frac{\partial V(\bar{v})}{\partial \bar{v}}+ F(\bar{v}) V(\bar{v}) = 0
\end{equation}
where $\kappa =\lambda/\alpha  $ with $\alpha = \frac{2\Delta}{(1+\Delta)}$ and
\begin{equation}\label{fvbaraa}
\begin{split}
F(\bar{v}) =& \frac{1}{2 
   \alpha^{2}   g(\bar{v})^4} \left(g(\bar{v})-\frac{(1-\Delta )}{4} 
   \right)^{2/(1+\Delta )}\\ &(\ell  (\ell +1) g(\bar{v})+ 2 \sigma \mu ).
\end{split}   
\end{equation}
In figure \eqref{fvbard12} we have shown the function $F(\bar{v})$  
for $\Delta=1/2$. 

To obtain some more information about the eigenvalue $\lambda$
we will first consider the behavior of the solutions to \eqref{kg4} around
$\bar{v}=0$. Expanding $V(\bar{v})$ around $\bar{v} \to 0$
\begin{equation}\label{seriesv}
V(\bar{v})=\bar{v}^s \sum_{n=0}^{\infty}{a_n \bar{v}^n}\quad 
F(\bar{v})=\sum_{n=0}^{\infty}{b_n \bar{v}^n}
\end{equation}
with $b_0\neq 0$ we obtain from \eqref{kg4} the indicial equation
\begin{equation}
s(s-\kappa)=0,
\end{equation}
which to leading order gives
\begin{equation}
V(\bar{v})= \alpha + \beta \bar{v}^{\kappa}
\end{equation}
and thus
\begin{equation}
\psi_\lambda = \alpha \bar{u}^{2\kappa/3} + \beta v^\kappa.
\end{equation}
Decomposing $\kappa=-i\omega + \epsilon$ into real and imaginary parts, 
we see that well-behaved solutions around $v=0$ require that 
$\epsilon\geq 0$. Note that this also means that around
$v=0$ the $\bar{u}$ dependent term is finite as $\bar{u}\ra 0$, 
in agreement with the results of the numerical integration presented
in the previous section. Obviously this implies a divergence for 
large $\bar{u}$, but our physical setup does not include this region.

To obtain further information about the global structure of the solutions 
to the wave-equation we can expand around large positive $\bar{v}$. 
For large $\bar{v}$ approaching ${\mathcal I}_+$ we make 
the substitution $\bar{v} = e^{\bar{x}}$ and to leading order we also have
$F(\bar{v})\sim c~\ell(\ell+1)/\bar{v}^{5/2}$, for some constant $c$. 
Together with the above substitution we obtain the equation
\begin{equation}
\ddot{V} - \kappa\dot{V} + c~\ell(\ell+1)e^{-5\bar{x}/2} =0.
\end{equation}
The leading large $\bar{x}$ solution is 
\be
V(x) = \gamma  + \delta e^{\kappa\bar{x}}
\ee
leading to (with $v=e^x$),
\begin{equation}\label{largev}
\psi_\lambda = \gamma \bar{u}^{2\kappa/3} + \delta e^{\kappa x} 
\end{equation}
and thus one has an outgoing wave of frequency $\omega$ 
for $\kappa = -i\omega$, requiring again that $\epsilon=0$. 
Note that the expansion around infinity has 
the same leading behavior as that around $v=0$ due to the fact that the 
non-derivative term in the differential equation is sub-leading in both 
cases. 

As in scattering problems for static space-times also here there will 
be a non-trivial linear relation
between the coefficients $\alpha,\beta$ of the expansion around
$v=0$ and the coefficients $\gamma,\delta$ of the expansion around
$v\ra\infty$. For square integrability of the outgoing waves at $\infty$ 
we require that $\gamma=0$ and thus the coefficients $\alpha$ and $\beta$
will then be fixed uniquely by this transformation. Note that $\gamma=0$
also guarantees that one has purely outgoing perturbations on ${\mathcal I}^+$.
The derivation of this transformation is beyond the scope
of the current article as the numerical errors do not allow a complete
and accurate integration from $\bar{v}=0$ all the way to $\bar{v}\ra\infty$.

As a consequence of the decomposition of the wave-function we can conclude
that none of the exact solutions $\psi_\lambda(u,v)$ with Gaussian initial
conditions can contain constant large $v$ components even though we found 
such behavior in the numerical integration. Writing the complete solution as
\be
\Psi(\bar{u},\bar{v}) = \int_{-\infty}^\infty \dd\omega a_\omega\bar{u}^{-i\omega}
\psi_\omega(\bar{v})
\ee
we can see that for large $v=e^x$ the wave-function is independent
of $\bar{u}$ and has the free wave-form
\be
\Psi(\bar{u},\bar{v}) \sim \int_{-\infty}^\infty \dd\omega a_\omega e^{-i\omega x}.
\ee
A gaussian profile in $v$ at some $u = u_0$ will continue to have 
an exponential fall-off for large $v$ for all $\bar{u}$ and thus
there is no possibility for a constant mode to develop during the evolution
in $\bar{u}$. 

To verify the deductions that follow from the above expansions we also
carried out the numerical integration of the differential equation 
for $\Delta=1/2$. To do this we take the explicit expression for $g(\bar{v})$ 
when $\Delta=1/2$ from \cite{OLoughlin:2013aa},
\begin {equation}\label {gv1/2}
\begin{split}
g (\bar {v}) _ {1/2} =& \frac{1}{8} \left(3+\frac{4}{3^{2/3}} \left(\sqrt[3]
{9 \bar{v}^3-\sqrt{3} \sqrt{\bar{v}^6 (27-64
   \bar{v}^3)}}\right.\right.\\     
&  \left. \left. +\sqrt[3]{9 \bar{v}^3+\sqrt{3} \sqrt{\bar{v}^6
   (27-64 \bar{v}^3)}}\right)\right).
\end{split}   
\end {equation}
We then used the NDSolve package in Mathematica to solve \eqref{kg4}.

\begin{figure}[h!]
        \centering
        \begin{subfigure}[b]{.4\textwidth}
                \includegraphics[width=\textwidth]{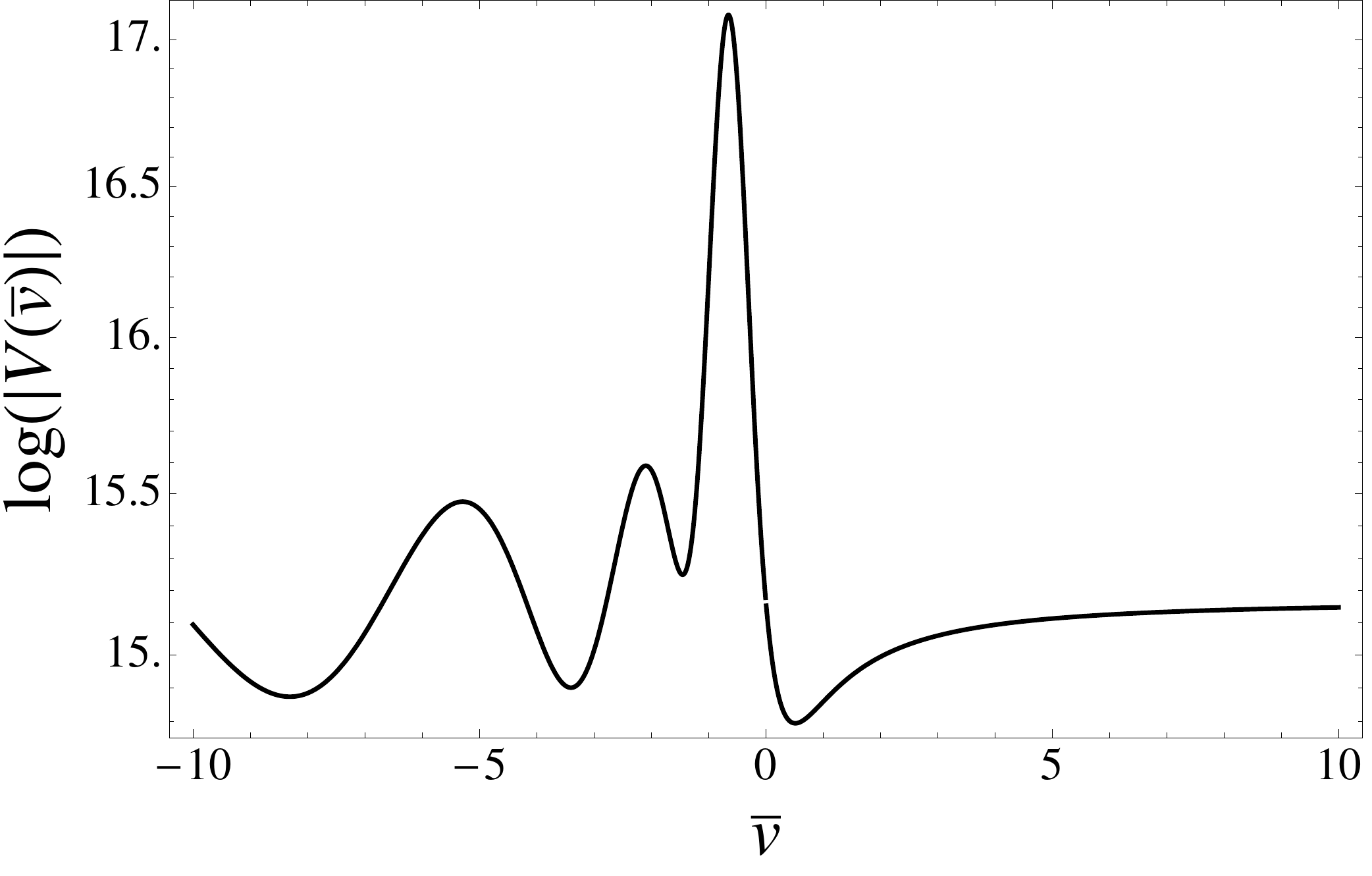}
                \caption{ $\sigma=0$, $\ell=1$ and $\kappa=7i$}
                \label{qnm12fw01}             
        \end{subfigure}\hfill%
               \begin{subfigure}[b]{.4\textwidth}
                \includegraphics[width=\textwidth]{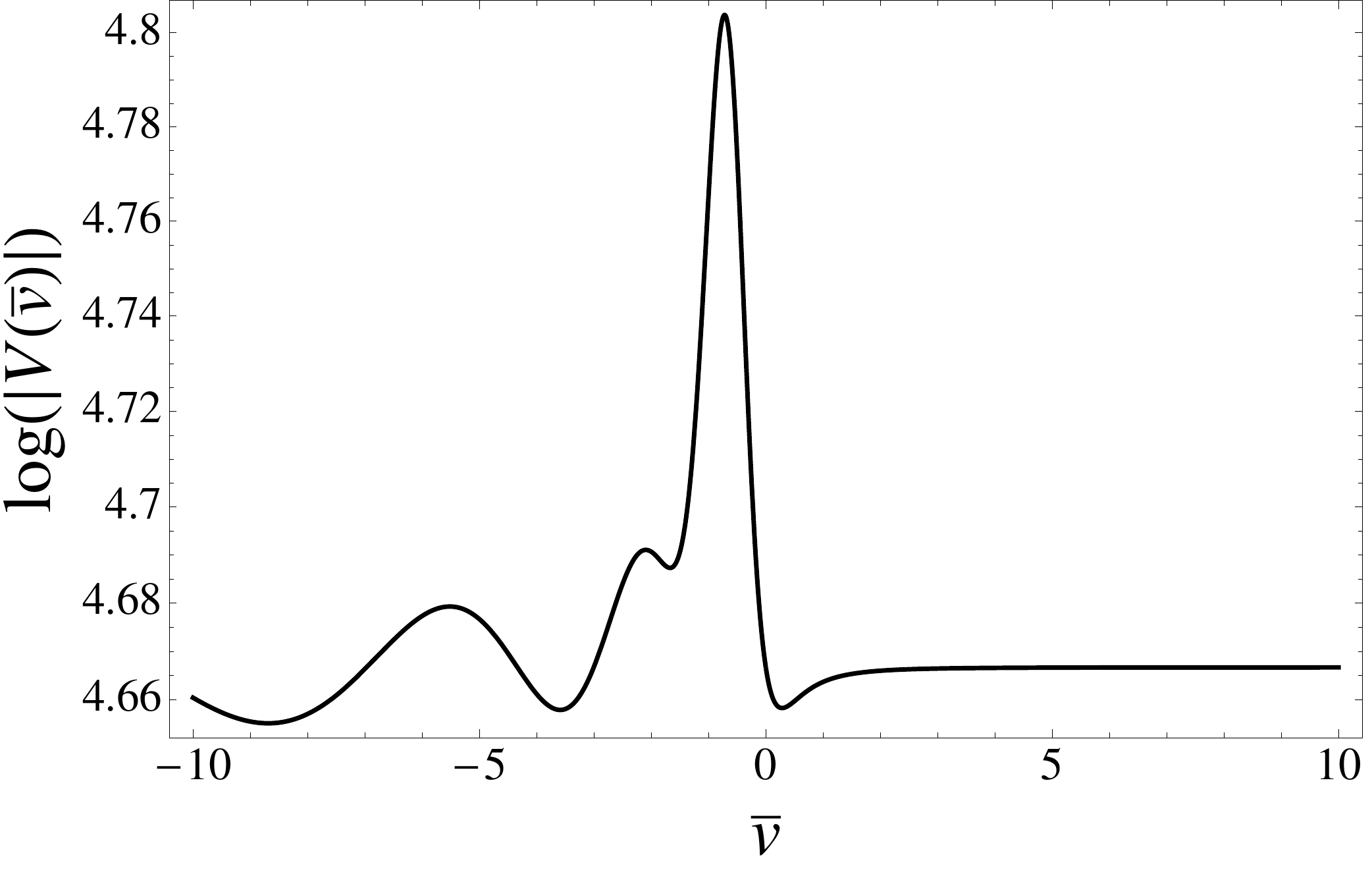}
                \caption{$\sigma=1$, $\ell=0$ and $\kappa=7i$}
                \label{qnm12bc01}
        \end{subfigure}\hfill%
       \begin{subfigure}[b]{.4\textwidth}
                \includegraphics[width=\textwidth]{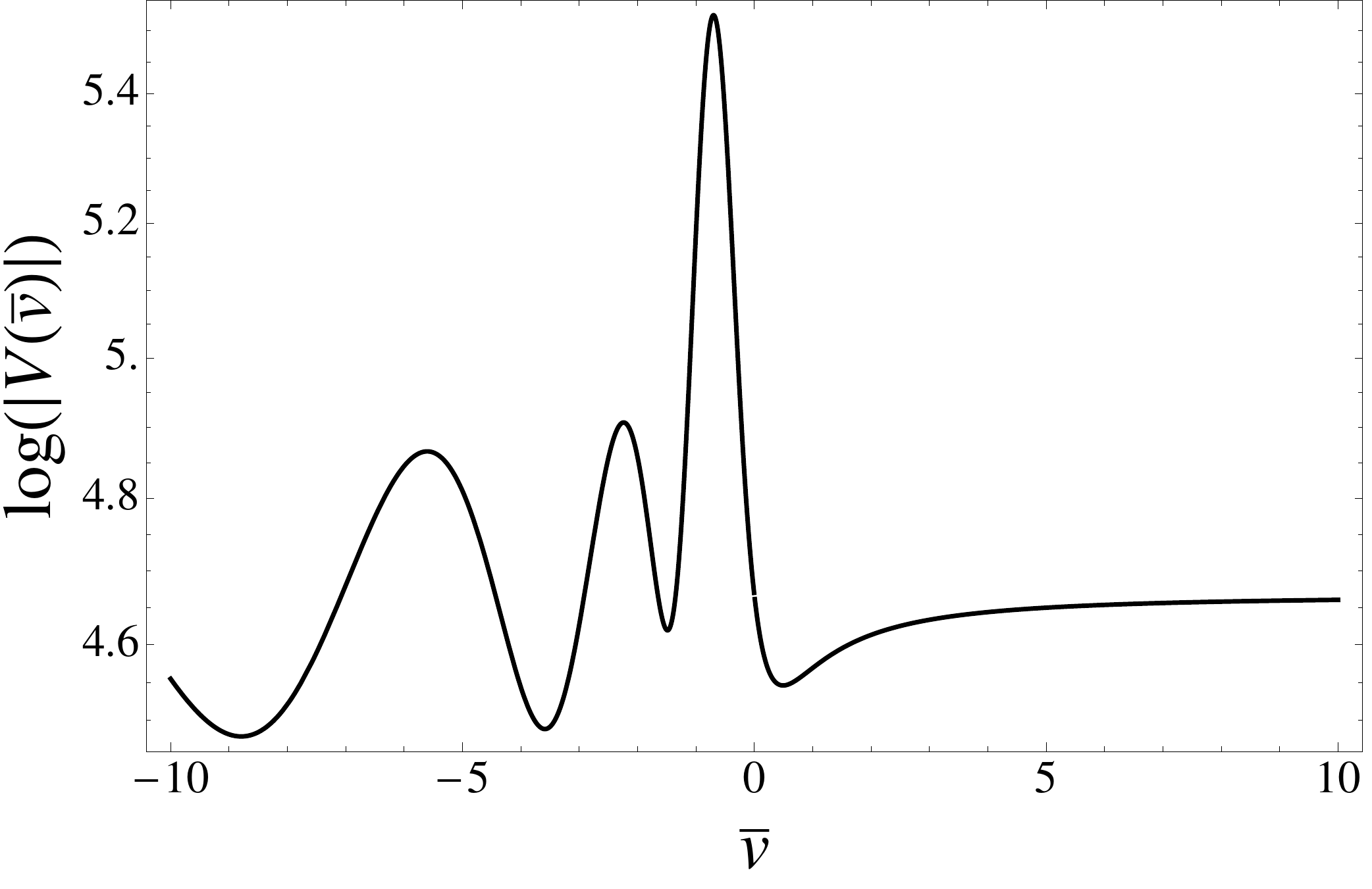}
                \caption{$\sigma=1$, $\ell=1$ and $\kappa=7i$}
                \label{qnm12together01}
        \end{subfigure}\hfill%
        \caption{Profile of electromagnetic, $\sigma=0,\, 1$ for $\ell=1$, and scalar perturbations, for $\ell=0$ and $1$, for $\Delta=1/2$ and $\epsilon=0$ .}\label{qnme0}
\end{figure}

\begin{figure}[h!]
\center{\includegraphics[width=7.5cm]{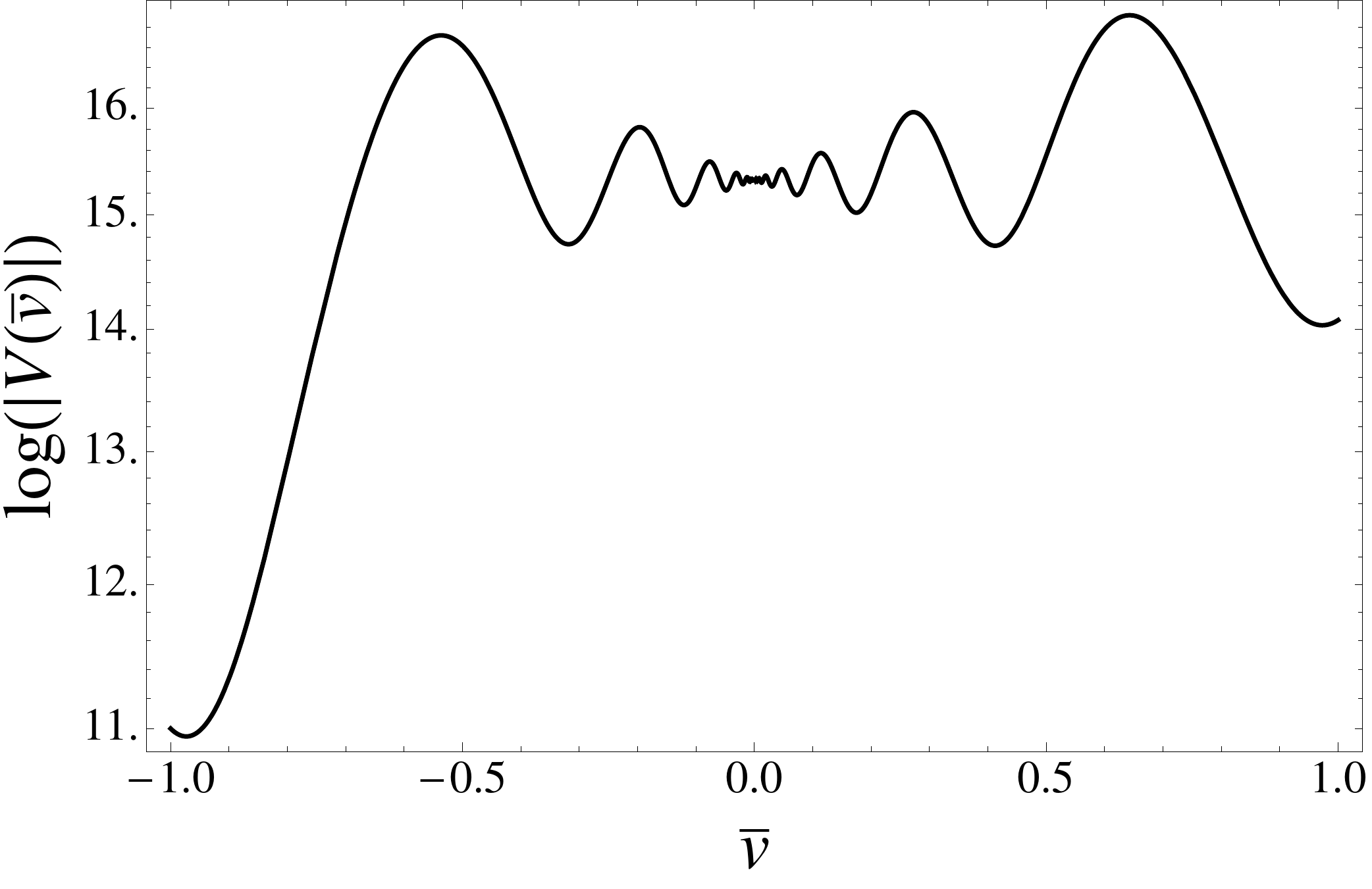}}
\caption{An example of the profile of an electromagnetic 
perturbation with $\sigma=0$, $\ell=1$ and $\kappa=-4i+1$}
\label{121}
\end{figure}

We carried out the integration in the following manner.
Due to the possible presence of singularities in the numerical 
integration through $\bar{v}=0$  we imposed initial conditions at two different 
points $\bar{v}=-0.000001$ and  $\bar{v}=0.000001$ and integrated forwards 
and backwards in $\bar{v}$ and to check the numerical stability we also 
carried out this calculation for smaller values of $|\bar{v}|$ with similar
results. The numerical solutions to these 
equations are presented in the figures \eqref{qnme0} and \eqref{121}. 
We show the solutions for $\epsilon = 0$ and also an example of 
a solution with $\epsilon =1$. 
Note in particular that the $\epsilon=0$ solutions show a ringing with 
variable frequency for $\bar{v}<0$ together with no oscillations for 
$\bar{v}>0$. This provides a confirmation of the ringing that was 
found in the previous section from the integration of the full 
wave equation for Gaussian initial conditions. 

The solutions with $\epsilon >0$ do not play a role in the evolution of 
initially analytic ingoing perturbations, but they may play a role in a
more complete analysis of QN like modes, as such modes arise
when one imposes boundary conditions such that there are no ingoing 
modes at ${\mathcal I}^-$.

\section{Summary and Comments}

We have provided further evidence for the presence of properties of 
scalar and electromagnetic fields/perturbations in the outgoing Vaidya
space time that support the hypothesis that this metric may provide a realistic semi-classical model for the end point of black hole 
evaporation. In particular, by the use of a decomposition of the wave-function
suggested by the presence of a homothety symmetry in the linear
mass Vaidya metric, we have reduced the spherically symmetric wave-equation
to an ODE. Using a mixture of analytic and numerical methods we have provided
strong evidence to support the hypothesis of the presence of QN 
like oscillations around the end-point of evaporation.

We have also shown that the normalisable modes exhibit oscillations as they 
approach $\bar{u}=0$ in both the solutions to the full PDE as 
well as in the individual modes obtained after separation. 

Although our analysis has a different focus to that of \cite{Nolan,Nolan:2006pz}
our results for the stability of the wave-equations on 
out-going Vaidya are in agreement with their results for the 
wave-equations on ingoing linear mass Vaidya. 

The biggest obstacle to further progress is the difficulty in the 
numerical calculation of the transformations required to propagate 
solutions about $v=0$ to $v=\infty$ which would provide more 
complete information about the modes $V_\lambda$. One possible
approach to this question is the large-D limit. As there exists
a Vaidya-metric in any dimension \cite{iyer1989vaidya}, one can 
take the large-D limit \cite{emparan,Emparan:2014cia} and thus obtain a 
simplification of the potential $F(\bar{v})$. One may then use this to 
obtain a WKB matching of $V_\lambda$ between the $\bar{v}=0$ expansion and 
that at $\bar{v}\ra\infty$, preliminary work is presented in \cite{thesis}. 

\bibliographystyle{apsrev4-1} 
\bibliography{xampl}

\begin{thebibliography}{9}

\bibitem{vaidya}
{\sc P.C. Vaidya}.
\newblock {\bf The external field of a radiating star in general relativity}.
\newblock {\em Current Sience}, {\bf 12 (06)}:183, 1943.

\bibitem{waugh862}
B. Waugh and Kayll Lake, {\bf Backscattered radiation in the Vaidya metric 
near zero mass},  Phys. Lett. A {\bf 116} (1986) 154.

\bibitem{bicak97}
J. Bicak and K.V. Kuchar, {\bf Null dust in canonical gravity},
Phys. Rev. D {\bf 56} (1997) 4878, arXiv:gr-qc/9704053.

\bibitem{hiscock82}
W.A. Hiscock, L.G. Williams and D.M. Eardley,
{\bf Creation Of Particles By Shell Focusing Singularities},
Phys. Rev. D {\bf 26} (1982) 751.

\bibitem{kuroda}
Y. Kuroda, {\bf Vaidya Space-time As An Evaporating Black Hole},
Prog. Theor. Phys. {\bf 71} (1984) 1422.

\bibitem{balbinot}
R. Balbinot, {\bf The back reaction and the small-mass regime},
Phys.\ Rev.\ D {\bf 33}  (1986) 1611–1615.

\bibitem{abdalla}
E. Abdalla, C.B.M.H. Chirenti and A. Saa, {\bf Quasinormal modes for the Vaidya metric}, Phys.Rev. D 
{\bf 74} (2006) 084029, arXiv:gr-qc/0609036.

\bibitem{bicak03}
J. Bicak and P. Hajicek, {\bf Canonical theory of spherically symmetric 
space-times with cross streaming null dusts}, 
Phys. Rev. D {\bf 68} (2003) 104016, arXiv:gr-qc/0308013.

\bibitem{ghosh}
S.G. Ghosh and N. Dadhich, {\bf On naked singularities in higher-dimensional 
Vaidya space-times}, Phys. Rev. D {\bf 64} (2001) 047501, arXiv:gr-qc/0105085.

\bibitem{harko}
T. Harko, {\bf Gravitational collapse of a hagedorn fluid in Vaidya geometry},
Phys. Rev. D {\bf 68} (2003) 064005, arXiv:gr-qc/0307064.

\bibitem{girotto}
F. Girotto and A. Saa, {\bf Semi-analytical approach for the Vaidya metric 
in double-null coordinates}, Phys. Rev. D {\bf 70} (2004) 084014, 
arXiv:gr-qc/0406067.

\bibitem{kawai}
H. Kawai, Y. Matsuo and Y. Yokokura, {\bf A Self-consistent Model of the Black 
Hole Evaporation}, Int. J. Mod. Phys. A {\bf 28} (2013) 1350050, 
arXiv:1302.4733 [hep-th].

\bibitem{fayos10}
F. Fayos and R. Torres, {\bf Local behaviour of evaporating stars and black 
holes around the total evaporation event}, 
Class. Quant. Grav. {\bf 27} (2010) 125011.

\bibitem{farley}
A.N.St. J. Farley and P.D. D'Eath, {\bf Vaidya space-time in black-hole 
evaporation}, Gen. Rel. Grav. {\bf 38} (2006) 425, arXiv:gr-qc/0510040.

\bibitem{OLoughlin:2013aa}
{\sc M. O'Loughlin}.
\newblock \href{http://arxiv.org/abs/1312.4702}{{\bf A linear mass Vaidya
  metric at the end of black hole evaporation}}.
\newblock {\em Phys. Rev.}, {\bf D91}(044020), 2 2015. 

\bibitem{qnmbs1}
{\sc K.D. Kokkotas and B.G. Schmidt}
\newblock {\bf Quasinormal modes of stars and black holes}.
\newblock {\em Living Rev. Rel.}, {\bf 2}: 2, 1999.

\bibitem{qnmbs2}
{\sc R. A. Konoplya, A. Zhidenko}.
\newblock {\bf Quasinormal modes of black holes: from astrophysics to string theory}.
\newblock {\em Rev.Mod.Phys.}, {\bf 83}: 793-836, 2011.

\bibitem{Hod:2002gb}
{\sc S. Hod}.
\newblock {\bf Wave tails in time dependent backgrounds}
\newblock {\em Phys. Rev.}, {\bf D66} 024001, 2002. 

\bibitem{Xue:2003vs}
{\sc L.H. Xue, Z.X. Shen, B. Wang and R.K. Su}.
\newblock {\bf Numerical simulation of quasi-normal modes in time dependent 
background}
\newblock {\em Mod. Phys. Lett.}, {\bf A19} 239, 2004. 

\bibitem{Shao:2004ws}
{\sc C.G. Shao, B. Wang, E.Abdalla and R.K. Su}. 
\newblock {\bf Quasinormal modes in time-dependent black hole background.}
\newblock {\em Phys. Rev. D}, {\bf 71} 044003, 2005. 

\bibitem{invai1}
{\sc F. Girotto and A. Saa}. 
\newblock {\bf Semi-analytical approach for the Vaidya metric in double-null coordinates.}
 \newblock {\em Phys. Rev. D},  {\bf 70} 084014, 2004. 

\bibitem{invai2}
{\sc E. Abdalla, C. B. M. H. Chirenti, and A. Saa}. 
\newblock {\bf Quasi-normal modes for the Vaidya metric.}
\newblock {\em Phys. Rev. D}, {\bf 74} 084029, 2006. 

\bibitem{price1972}
{\sc R. H. Price}.
\newblock {\bf Nonspherical perturbations of relativistic gravitational collapse. I. Scalar and gravitational perturbations.}
 \newblock {\em Phys. Rev. D}, {\bf 5:} 2419 - 2438, 1972.
 
 \bibitem{Hod:2009my}
{\sc Hod, Shahar}.
\newblock {\bf How pure is the tail of gravitational collapse?}
\newblock {\em Class. Quant. Grav.}, {\bf 26} 028001, 2009. 

\bibitem{Waugh:1986aa}
{\sc B.~Waugh and K. Lake}.
\newblock {\bf Double-null coordinates for the Vaidya metric}.
\newblock {\em Phys. Rev.}, {\bf D34}(10):2978--2984, 1986.

\bibitem{Zannias:1990aa}
{\sc T.~Zannias}.
\newblock {\bf Spacetimes admitting a three-parameter group of isometries and
  quasilocal gravitational mass}.
\newblock {\em Phys. Rev.}, {\bf D41}(10):3252--3254, 1990.

\bibitem{Nielsen:2008kd}
{\sc A.~B. Nielsen and D.-han Yeom}.
\newblock {\bf Spherically symmetric trapping horizons, Misner-Sharp mass and
  black hole evaporation}.
\newblock {\em Int. J. Mod. Phys.}, {\bf A24}:5261--5285, 2009.

\bibitem{Gao:2005yq}
{\sc S. Gao and J.~P.S. Lemos}.
\newblock {\bf {The Tolman-Bondi-Vaidya spacetime: Matching timelike dust to
  null dust}}.
\newblock {\em Phys.Rev.}, {\bf D71}:084022, 2005.

\bibitem{Hiscock:1982pa}
{\sc W.A. Hiscock, L.G. Williams, and D.M. Eardley}.
\newblock {\bf {Creation Of particles by shell focusing singularities }}.
\newblock {\em Phys.Rev.}, {\bf D26}:751--760, 1982.

\bibitem{gvs}
{\sc A. Wang and Y. Wu}.
\newblock {\bf Generalized Vaidya solutions}.
\newblock {\em General Relativity and Gravitation}, {\bf 31}(1):107--114, 1999.

\bibitem{Geroch:1968ut}
{\sc R.~P. Geroch}.
\newblock {\bf {What is a singularity in general relativity?}}
\newblock {\em Annals Phys.}, {\bf 48}:526--540, 1968.

\bibitem{Lake:1991aa}
{\sc K. Lake}.
\newblock {\bf Naked singularities in gravitational collapse which is not
  self-similar}.
\newblock {\em Phys. Rev.}, {\bf D43}(4):1416--1417, 1991.

\bibitem{Unruh:1985aa}
{\sc W.~G. Unruh}.
\newblock {\bf Collapse of radiating fluid spheres and cosmic censorship}.
\newblock {\em Phys. Rev.}, {\bf D31}(10):2693--2694, 1985.

\bibitem{ReggeWheeler}
{\sc T.~Regge and J.~A.~Wheeler}
\newblock{\bf Stability of a Schwarzschild singularity}
\newblock{\em Phys.Rev.} {\bf 108} 1063, 1957.

\bibitem{Gundlach:1994}
{\sc P. Gundlach and Pullin}.
\newblock {\bf Late-time behavior of stellar collapse and explosions:
I. Linearized perturbations}
\newblock {\em Phys. Rev.}, {\bf D49} 883, 1994. 

\bibitem{Nolan}
{\sc B.C. Nolan and T.J. Waters}
\newblock {\bf Even perturbations of self-similar Vaidya space-time}
\newblock {\em Phys. Rev.}, {\bf D71} 104030, 2005.

\bibitem{Nolan:2006pz}
{\sc B.C. Nolan. }
\newblock {\bf Odd-parity perturbations of self-similar Vaidya space-time}
\newblock {\em Class. Quant. Grav.}, {\bf 24} 177-200, 2007.


\bibitem{iyer1989vaidya}
{\sc Iyer, BR and Vishveshwara, CV}
\newblock {\bf The Vaidya solution in higher dimensions}
 \newblock {\em Pramana-Journal of Physics}, {\bf 32}, 6, 749-752, 1989.

\bibitem{Emparan:2014cia}
{\sc R.Emparan, and K.Tanabe}.
\newblock {\bf Universal quasi-normal modes of large D black holes.}
 \newblock {\em Phys. Rev. D}, {\bf 89} 6, 064028, 2014.
 
\bibitem{emparan}
{\sc R. Emparan and R. Suzuki and K. Tanabe}.
\newblock {\bf The large $D$ limit of general relativity.}
 \newblock {\em JHEP}, {\bf 1309} 009, 2013.

\bibitem{thesis}
{\sc S. Nafooshe}
\newblock {\bf Aspects of micro black hole evaporation}
\newblock {\em University of Nova Gorica}, Ph.D thesis, 2015

\end{thebibliography}

\end{document}